# Short Status Report Of The Nucleon Time-Like Form Factors Measurements

Alessandro De Falco, Università di Cagliari/INFN Sezione di Cagliari, Italy

*Abstract*

The measurements of the nucleon electromagnetic form factors in the time-like region are reviewed. Several unexpected features deserving new high precision, high statistics measurements are emphasised.

*Introduction*

The importance of a good experimental knowledge of the nucleon electromagnetic form factors is quite clear, since they parameterise the nucleon internal structure from low $q^2$, where they describe the nucleon charge distribution and magnetisation current, to high $q^2$, where they probe the valence quark distribution. QCD predictions from non-perturbative to perturbative regime can then be tested according to their capability to reproduce the form factors measurements for any value of the momentum transfer.

A big effort was done to measure the form factors in the space-like region, where the electric and magnetic form factors were determined independently, both for the proton and the neutron, in a wide $q^2$ range, and some general trend could be established.

In the time-like region the experimental situation is less clear. Few measurements with limited statistics are currently available, showing unexpected features that deserve more accurate studies.

In the following, the main expectations concerning the time-like form factors will be described and compared to the existing data.

*General properties of the time-like form factors*

The measured quantities are the Sachs form factors, that are linear combinations of the Dirac and Pauli ones:

$$G_E(q^2) = F_1(q^2) + \frac{q^2}{m^2} F_2(q^2); \quad (1)$$
$$G_M(q^2) = F_1(q^2) + F_2(q^2);$$

where $m$ is the nucleon mass, and the squared 4-momentum transfer $q^2$ is negative in the space-like region, where the form factors are real. In the time-like region ($q^2 > 0$), studied by means of the reaction $e^+e^- \leftrightarrow N\bar{N}$, the form factors are in general complex, and are assumed to be the analytical continuation of the space-like ones through infinity. It is then expected that asymptotically the space-like and time-like form factors must coincide, as also predicted by pQCD. Moreover, at the threshold for a nucleon-antinucleon pair production, $|G_E| = |G_M|$, as follows immediately from eq. 1. This relation is often assumed to hold also above threshold. From the differential cross section formula for the process $e^+e^- \to N\bar{N}$:

$$\frac{d\sigma}{d\Omega} = \frac{\alpha^2 \beta C}{4q^2} \left[ |G_M|^2 (1 + \cos^2\theta) + \frac{4m^2}{q^2} |G_E|^2 \sin^2\theta \right], \quad (2)$$

where $\alpha$ is the fine structure constant, $\theta$ and $\beta$ are the nucleon polar angle and velocity in the CM frame, and C is the Coulomb correction factor, it follows that at threshold the polar angle distribution must be flat, while at higher $q^2$ the contribution of the term proportional to $G_M$ dominates and the $\theta$ distribution approaches the form $1 + \cos^2\theta$. At intermediate values the two contributions can be disentangled only if the $\theta$ dependence of the cross section can be measured.

The ratio between neutron and proton form factor squared can be roughly estimated to be $\sim (q_d/q_u)^2 = 0.25$, assuming an interaction between pointlike isolated valence quarks. More refined calculations can give different results, but any model where the interaction is based on the valence quarks can hardly foresee a neutron form factor bigger than the proton one. It is interesting to observe that former descriptions based on the vector meson dominance [1] and Skyrme based models [2] give $|G_M^n| >= |G_M^p|$.

*Time-like proton form factor at low $q^2$*

The proton form factor was measured near threshold with best precision by the PS170 experiment, that used the LEAR facility at CERN to study the reaction $\bar{p}p \to e^+e^-$ [3]. A sample of more than 3000 events was obtained, with an estimated background of $\sim 1\%$. The form factor, plotted in fig. 1 together with other measurements [4] under the hypothesis $G_E = G_M$, shows a steep decrease just above threshold, followed by a flat trend for higher energies. A possible interpretation will be discussed later.

The hypothesis that the magnetic and electric form factor were equal was tested studying the polar angle distribution at different energies. The measurements are compatible with a flat distribution, indicating the equality of the two form factors. However, when leaving the ratio $|G_E|/|G_M|$ as a free parameter of the fit, data seem to prefer a decrease in that ratio with the energy, even if no firm conclusion can be stated due to the limited statistical accuracy.

*Time-like proton form factor at high $q^2$*

The measurement at higher $q^2$ was first performed by the E760 experiment [5] at Fermilab and improved by its upgrade E835 [6]. The experiment was designed to study the charmonium production in $\bar{p}p$ collisions through its decay in electromagnetic final states. Despite of the low cross section, the high luminosity and the capability of the detector to determine clearly the $e^+e^-$ pairs on the top of a high hadronic background made the measurement possible. The final sample from E835 consists of 144 events, with a background lower than one event per point. The results, under the $|G_E| = |G_M|$ hypothesis, are plotted in fig. 2, where the dashed line represents the pQCD fit at large momentum transfers. For comparison, the dipole behaviour of the form factors in the space-like region is also plotted (dot-dashed line). It can be noticed that, while the pQDC fit reproduces the data already at $q^2 = 4\,\text{GeV}^2$, the $|G_M|$ values are a factor of two bigger than the corresponding ones in the space-like region. It has to be stressed that calculations based on the quark-diquark model of the proton can describe these data [7], assuming that the discrepancy will disappear at higher $q^2$.

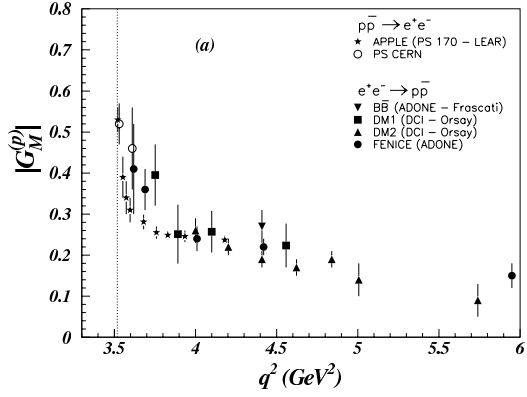

Figure 1: Proton form factor at low $q^2$. The dashed line indicates the $p\bar{p}$ threshold.

## Time-like neutron form factors

The neutron form factors were measured by the FENICE collaboration [8], using the ADONE $e^+e^-$ collider in Frascati. The apparatus consisted of a non magnetic calorimeter with wide angular coverage. The channel $e^+e^- \to n\bar{n}$ was studied from threshold to $q^2 \sim 6\,\text{GeV}^2$ by detecting the antineutron by means of the characteristic star pattern given by its annihilation products, and measuring its time of flight. The neutron signal was not required, due to its low detection efficiency.

A final sample of 74 events, corresponding to an integrated luminosity of $0.4\,\text{pb}^{-1}$, was obtained. In fig. 3, the neutron form factor measured by FENICE is displayed together with two points from DM2, one coming from a direct measurement [9], and the other estimated from the $\Lambda$ form factor on the basis of U-spin invariance considerations [10]. As the differential cross section measured by FENICE seems to prefer a $1 + \cos^2\theta$ dependence on the polar angle, the results are given under the hypothesis $|G_E| = 0$. However, the limited statistics does not allow to draw any final statement about this assumption.

According to these measurements, $|G_M^n| > |G_M^p|$ over the entire energy range covered by the experiment, while the pQCD calculations expect a proton form factor about twice the neutron one.

The point closest to threshold deserves further considerations, since the uncertainty on the effective collider energy, of some MeV, plays an important role for this point. If a shift of few MeV is assumed, $|G_M|$ near threshold stays roughly constant. Otherwise, if the nominal value of the CM energy is assumed to be correct, no signal is observed: this behaviour is consistent with the hypothesis $|G_E| = 0$, since at threshold the two form factors must be equal. A definite conclusion cannot be drawn on the basis of the FENICE data.

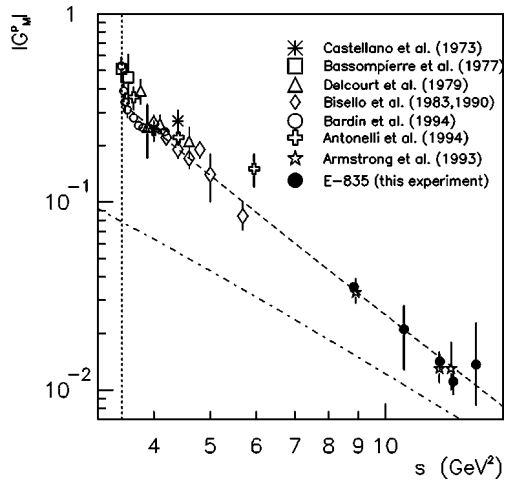

Figure 2: Proton form factor including data at high $q^2$.

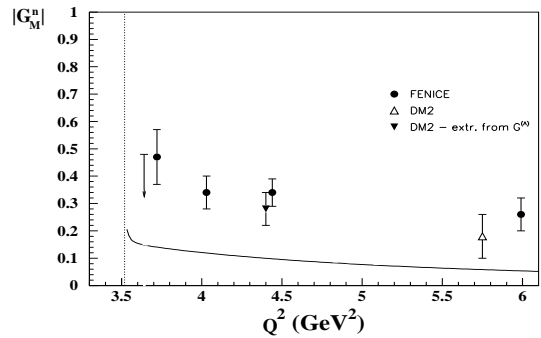

Figure 3: Neutron form factor. The solid line represents the parameterisation of the proton form factor.

*Multihadronic cross section*

The FENICE experiment performed also the measurement of the total $e^+e^- \to$ hadrons cross section [11], as shown in fig. 4 where the average values over previous experiments is also plotted. A dip in the cross section just below the threshold for the production of a nucleon-antinucleon pair is seen in the data. Similar structures had been observed by DM2 in the $e^+e^- \to 3\pi^+3\pi^-$ channel, and in diffractive photoproduction of $3\pi^+3\pi^-$ by The Fermilab experiment E687. The dip may be justified, together with the steep threshold behaviour of the proton form factor, by the presence of a narrow vector meson resonance under threshold, weakly coupled to the $e^+e^-$ and multihadronic channels. The curves plotted in fig. 4 represent the results of the fit to the multihadronic cross section and to the proton form factor, assuming the presence of such a resonance, and lead to a mass $M \sim 1.87$ GeV and a width $\Gamma \sim 10 - 20$ MeV. This state is consistent with an $N\bar{N}$ bound state, the so-called baryonium.

More recently, the BES [12] collaboration has found a narrow enhancement near threshold in the invariant $p\bar{p}$ mass spectrum from radiative $J/\psi \to \gamma p\bar{p}$ decays, consistent with a resonance with quantum numbers either $J^{PC} = 0^{-+}$ or $J^{PC} = 0^{++}$. The resonance mass peak is below the $M_{p\bar{p}} = 2m_p$ threshold and its width is $\Gamma < 30$ MeV.

It is also interesting to observe that a similar enhancement is seen in the $p\bar{p}$ invariant mass distributions near $M_{p\bar{p}} = 2m_p$ in the $B^+ \to K^+ p\bar{p}$ [13] and $B^0 \to D^0 p\bar{p}$ [14] decays by the Belle experiment.

*Conclusions*

Several unexpected features have been observed in the measurements of the time-like electromagnetic nucleon form factors:

- the neutron magnetic form factor is bigger than the proton one in the explored energy range ($q^2 < 6$ GeV$^2$);

- the differential cross section measurement suggests that the neutron electric form factor is negligible already near threshold;

- the steep decrease of the proton form factor near threshold, and the presence of a dip in the $e^+e^- \to$ hadrons in the same region, suggest the presence of a narrow resonance just below threshold.

The accuracy of present data is not good enough to draw firm conclusions. High statistics measurements with a wide angular coverage are needed to disentangle the contributions of the two form factors both for the proton and the neutron.

The author wishes to thank R. Baldini for fruitful discussions.

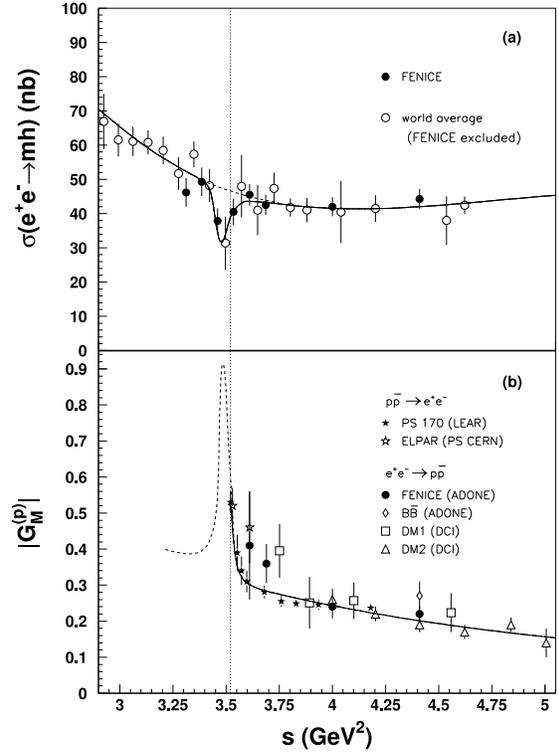

Figure 4: Top: total hadronic cross section superimposed to the results of the fit with a narrow resonance under threshold. Bottom: comparison of the proton form factor data to the expected behaviour for the presence of such a resonance.